# Del-Net: A Single-Stage Network for Mobile Camera ISP

Saumya Gupta[†], Diplav Srivastava[†], Umang Chaturvedi[†], Anurag Jain, Gaurav Khandelwal

Samsung R&D Institute - Bangalore, India
{saumya.gupta, diplav.s, u.chaturvedi, anuragm.jain, gaurav.k7}@samsung.com

## ABSTRACT

The quality of images captured by smartphones is an important specification since smartphones are becoming ubiquitous as primary capturing devices. The traditional image signal processing (ISP) pipeline in a smartphone camera consists of several image processing steps performed sequentially to reconstruct a high quality sRGB image from the raw sensor data. These steps consist of demosaicing, denoising, white balancing, gamma correction, colour enhancement, etc. Since each of them are performed sequentially using hand-crafted algorithms, the residual error from each processing module accumulates in the final reconstructed signal. Thus, the traditional ISP pipeline has limited reconstruction quality in terms of generalizability across different lighting conditions and associated noise levels while capturing the image. Deep learning methods using convolutional neural networks (CNN) have become popular in solving many image-related tasks such as image denoising, contrast enhancement, super resolution, deblurring, etc. Furthermore, recent approaches for the RAW-sRGB conversion using deep learning methods have also been published, however, their immense complexity in terms of their memory requirement and number of Mult-Adds[*] make them unsuitable for mobile camera ISP. In this paper we propose Del-Net – a single end-to-end deep learning model – to learn the entire ISP pipeline within reasonable complexity for smartphone deployment. Del-Net is a multi-scale architecture that uses spatial and channel attention to capture global features like colour, as well as a series of lightweight modified residual attention blocks to help with denoising. For validation, we provide results to show the proposed Del-Net achieves compelling reconstruction quality. It has a competitive trade-off between visual quality and complexity, with a ~90% reduction in Mult-Adds as compared to the state-of-the-art networks.

---

[†] Authors contributed equally
[*] Mult-Adds is the number of composite multiply-accumulate operations for an image.

## 1 INTRODUCTION

The camera in smartphones has evolved from being a mere feature to being the most common method of photography, and have become a major selling point for smartphones. This has prompted manufacturers to make significant investments in technology, resulting in remarkable improvements in both hardware and the ISP pipeline, and thus pushing the limit of smartphone imaging capability. As a result, image quality of smartphones have now become comparable to that of high-end DSLR and dedicated cameras.

The traditional image signal processing pipeline in a smartphone camera consists of several software-based image processing steps which are performed sequentially to reconstruct a high quality sRGB image from the raw sensor data. The steps involved in reconstructing a high quality sRGB image are demosaicing, denoising, gamma correction, colour correction, etc. Since each of them are performed sequentially using hand-crafted algorithms, the residual error from each processing module accumulates in the final reconstructed signal. Even a small amount of change in a parameter associated with one of the modules results in an entirely different reconstructed image. Consequently, it is time consuming to tune each of these modules to obtain the high quality image. Moreover, in the traditional ISP, an individual module cannot recover the signal loss resulting from previous modules nor can it control the output of the previous modules. Thus, the traditional ISP pipeline has limited reconstruction quality in terms of generalizability across different lighting conditions and associated noise levels while capturing an image. Mobile devices also deal with hardware limitations due to their compact size.

In this paper, we propose Del-Net – a single end-to-end deep learning model – to learn the entire ISP pipeline to convert raw bayer data to high quality sRGB image within reasonable complexity suitable for smartphone deployment. Del-Net is a multi-scale architecture that uses a combination of Spatial and Channel Attention (SCA) blocks [1] and Enhancement Attention Modules (EAM) blocks [2]. SCA blocks help to capture global details both channel-wise and spatial-wise while EAM blocks help in denoising. In this paper, we present results on the Zurich dataset [3], which is a raw-image to RGB-image dataset. The results show a competitive trade-off between visual quality and complexity, with a ~90% reduction in Mult-Adds as compared to the state-of-the-art networks PyNET [4], AWNet [5], and MW-ISPNet [6]. We thus claim that Del-Net is suitable for smartphone deployment. We further provide an ablation study to justify the presence of the individual blocks in our ISP pipeline.

In summary, our contributions are:

- To explore the effectiveness and performance boost on combining single-scale and multi-scale architectures, along with the attention mechanism.
- We propose Del-Net which generates images visually comparable to the state of the art networks, along with a significant reduction in Mult-Adds, therefore making it ideal for smartphone deployment.

## 2  RELATED WORK

In this section, we first discuss the existing learnable algorithms for individual modules in a traditional ISP, followed by a brief discussion on the existing learnable algorithms for an end-to-end ISP pipeline.

### 2.1 Traditional ISP

The steps involved in a traditional ISP are demosaicing, denoising, white balancing, colour correction, lens shading correction, defect pixel correction, gamma correction, local tone mapping, auto exposure, auto focus, colour space conversion, and many more. Demosaicing techniques [7] involve converting a raw bayer input captured using a bayer colour filter array (CFA) into a colour RGB image. Denoising methods [8] aim at removing noise from the raw image. The process of white balancing [9] eliminates unnatural colour cast so that the object which appears white in person looks white in the image. Gamma correction methods [10] compensate for the nonlinearity of relative intensity as the frame buffer value changes in the output display. Tone mapping procedures [11] combine different exposures together in order to increase the local contrast within disparate regions of a high dynamic range (HDR) scene.

### 2.2 End-to-End ISP Pipeline

Recently it has been shown that the performance of a smartphone camera ISP can be significantly improved by the use of deep learning techniques. The AIM 2020 Challenge on Learned Image Signal Processing Pipeline [6] is a step in this direction where the participating teams had to solve a real-world RAW-to-sRGB mapping problem using the Zurich dataset for training. The authors of Camera-Net [12] split the ISP pipeline into two weakly correlated subtasks, namely restoration and enhancement, and separately address the two groups of subtasks. Andrey et al. [4] proposed a five level pyramidal CNN architecture PyNET to perform end-to-end ISP steps. AWNet [5] is a two-branch CNN network that utilizes the attention mechanism and wavelet transformation to tackle the learnable ISP problem. Zhu, Yu, et al. [13] proposed EEDNet, which uses Channel Attention Residual Dense Block (CARDB) block in skip connections and was trained using a novel ClipL1 loss [13]. MW-ISPNet [6] is an end-to-end ISP pipeline which consists of a multi-level wavelet ISP network and takes advantage of the MWCNN [14] and RCAN [15] architectures. However, each of the networks mentioned above are computationally heavy and are thus unsuitable for smartphone deployment. In this paper, we propose Del-Net – a light and effective network for end-to-end ISP pipeline which has a competitive trade-off between image quality and computation, and is suitable for smartphone deployment.

## 3  PROPOSED METHOD

This section describes the architecture of our deep learning image signal processing pipeline. The proposed model converts a raw bayer image to a sRGB image. This end-to-end transformation involves many complex tasks such as demosaicing, denoising, gamma correction, etc. To realize these tasks, we propose a deep learning model and its working can be formulated as:

$$I_F = F_{Del-Net}(I_0, \theta)$$

where $I_0 \in \mathbb{R}^{H \times W \times 1}$ is the input raw bayer image having resolution $H \times W$; $I_F \in \mathbb{R}^{H \times W \times 3}$ is the output sRGB image, and $F_{Del-Net}$ is our proposed Del-Net model parameterized by $\theta$.

### 3.1 Model Architecture

The proposed model consists of two sub-structures. The first is a flat structure made up of Enhancement Attention Modules (EAM) [2]. The second is a modified UNet [16], which contains Spatial and Channel Attention (SCA) blocks [1] (also termed as Dual Attention) as its processing blocks at each level of the encoder and the decoder. Since the use of the flat structure allows us to operate on single-scale resolution, we get spatially precise results but the overall semantic details are not satisfying. To overcome this, we integrate the UNet model into our design. The UNet structure increases the receptive field, which in turn results in semantically accurate results. Thus, we exploit the benefits of both the single-scale as well as the multi-scale structures. Fig. 1 shows Del-Net's architecture.

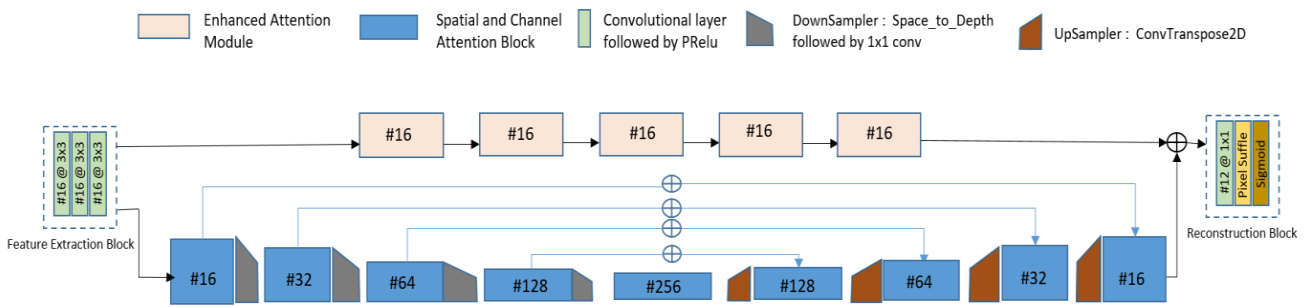

**Figure 1:** Architecture of the proposed Del-Net model, where #N denotes the number of channels outputted by that block/layer.

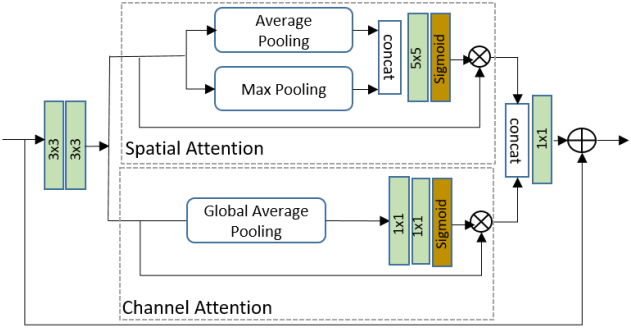

**Figure 2:** Schematic diagram of Spatial and Channel Attention Block

### 3.1.1 Spatial and Channel Attention Block

The schematic diagram of the SCA block is shown in Fig. 2. In the modified UNet sub-part of the model, we utilize attention mechanism spatially as well as per channel. The attention method helps in selecting features that are most useful while discarding the less relevant ones. We use skip connection in each SCA block so as to avoid saturation of learning of the features. On the input features, we first apply 3x3 convolution operations followed by PReLU [17] to perform feature transformation. We then apply attention mechanism on the newly constructed features. The channel attention makes use of global average pooling and sigmoid activation function to yield a feature descriptor (attention weight) for each input channel. Finally, the attention weights are multiplied with the transformed features. Similarly, spatial attention learns feature weight maps for inter-spatial dependencies. It utilizes combined results from global and max-pooling along the channel dimension. The combined map is then passed through convolution and sigmoid activation function to output weight maps. Finally, we combine the results of both the channel and the spatial attention modules. The input and output feature volumes for each SCA block are of the form of $H \times W \times N$, where $N$ is the number of channels used.

### 3.1.2 Enhancement Attention Modules

The structure of EAM utilizes both feature generation as well as feature attention blocks. The features are first transformed using convolutional layers followed by the PReLU activation function. On the input features, a series of convolutional layers are applied with different dilation rates in order to make use of varied receptive fields. The resultant features are then merged and passed through convolutional layers with local skip connection. At last, attention weights are learned for each channel to enhance the weights of the transformed features. The attention module is similar to channel attention in the SCA blocks. The schematic diagram of the EAM block is shown in Fig. 3. Each EAM block receives input of dimension $H \times W \times N$ where $N$ is the number of input channels and the output generated is in the same form of $H \times W \times N$.

## 3.2 Loss Function

Del-Net is trained with a loss function $L$ described below. $I$ is the ground truth while $\hat{I}$ is the Del-Net output.

$$L(I, \hat{I}) = \lambda_1 * L1_{modified} + \lambda_2 * L_{SSIM} + \lambda_3 * L_{Perceptual}$$
$$L1_{modified}(I, \hat{I}) = \|I - \hat{I}\|_1 + \|log(max(I, \varepsilon) - log(max(\hat{I}, \varepsilon))\|_1$$
$$L_{SSIM}(I, \hat{I}) = 1 - MS\text{-}SSIM(I, \hat{I})$$

where the terms used are defined as follows:
- $\varepsilon = 0.001$
- $\lambda_n$ are the weight terms determined empirically and specified in Table I
- MS-SSIM is as described in [18]
- $L_{Perceptual}$ is as described in [19]

$L1_{modified}$ and $L_{SSIM}$ help in achieving good trade-off between denoising and detail retention, while $L_{Perceptual}$ is for the refinement of colour and content appearance.

## 4 EXPERIMENTATION

### 4.1 Dataset

To enhance smartphone images, the Zurich dataset provides 48043 RAW-RGB image pairs (of size 448×448×1 and 448×448×3 respectively). The training data has been divided into 46,839 image pairs for training and 1,204 ones for testing. In addition, 168 full resolution image pairs are used for perceptual validation. For data pre-processing and augmentation, we normalize the input data and perform vertical and horizontal flipping.

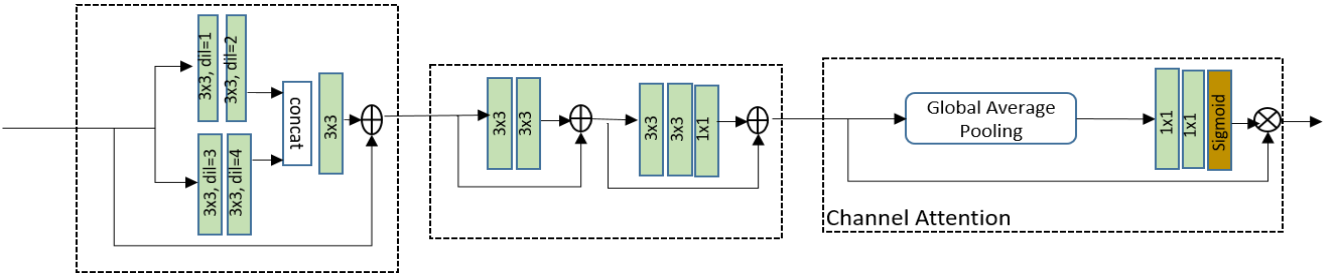

**Figure 3:** Schematic diagram of Enhancement Attention Modules

## 4.2 Training Details

Del-Net is trained using the PyTorch framework [20] and eight *NVIDIA TESLA P40 (24GB)* GPUs. The details regarding hyper-parameters used in training have been tabulated in Table I.

TABLE I. TRAINING CONFIGURATION

| Hyper parameters | Values |
|---|---|
| Input Size | 448×448×1 |
| Learning rate | $1\times10^{-4}$ |
| No. of epochs | 800 |
| Optimizer | AdamW [21] optimizer, with $\beta_1 = 0.9$, $\beta_2 = 0.999$ |
| Batch size | 64 |
| $\lambda_1$ | 0.85 |
| $\lambda_2$ | 0.15 |
| $\lambda_3$ | 1.0 |

## 5 RESULTS

We provide quantitative as well as qualitative results for Del-Net, and compare it with the results of PyNET, AWNet, and MW-ISPNet. For quantitative evaluation, we use the testset provided in the Zurich dataset which has 1,204 images. For qualitative evaluation, we use the full resolution testset in the Zurich dataset which contains 168 images. The quantitative and qualitative results for PyNET, AWNet, and MW-ISPNet have been generated using the model weights and codes shared by the authors. ※

## 5.1 Quantitative Evaluation

For quantitative evaluation, we compute the PSNR and SSIM which are the standard metrics used for comparing RAW-sRGB methods, as well as a colour fidelity loss CIE2000 [22]. Our solution Del-Net has a competitive trade-off between the accuracy metrics and complexity, as shown in Table II and Fig. 4. Furthermore, Del-Net has considerably fewer Mult-Adds and Params compared to PyNET, AWNet, and MW-ISPNet. Mult-Adds is the number of composite multiply-accumulate operations for a single image. Table II reports values of Mult-Adds for an input image of size 2976×4000×1 with output image size 2976×4000×3, which is equivalent to a 12MP image. Del-Net has 89.8%, 94.8%, and 97.4% fewer Mult-Adds compared to MW-ISPNet, AWNet, and PyNET respectively.

TABLE II. COMPARISON OVER AVERAGE PSNR, SSIM, CIE2000, MULT-ADDS, PARAMS

| Network | PSNR | SSIM | CIE2000 | Mult-Adds ($10^{12}$) | Params ($10^6$) |
|---|---|---|---|---|---|
| PyNET | 21.19 | 0.746 | 9.99 | 20.3 | 47.56 |
| AWNet | 21.86 | 0.780 | 9.76 | 10.2 | 96.06 |
| MW-ISPNet | **21.91** | **0.784** | 9.71 | 5.2 | 29.22 |
| Del-Net | 21.46 | 0.745 | **9.64** | **0.53** | **2.68** |

※ PyNet : https://github.com/aiff22/PyNET-PyTorch
AWNet : https://github.com/Charlie0215/AWNet-Attentive-Wavelet-Network-for-Image-ISP
MW-ISPNet : https://github.com/ZZL-1998/MW-ISPNet

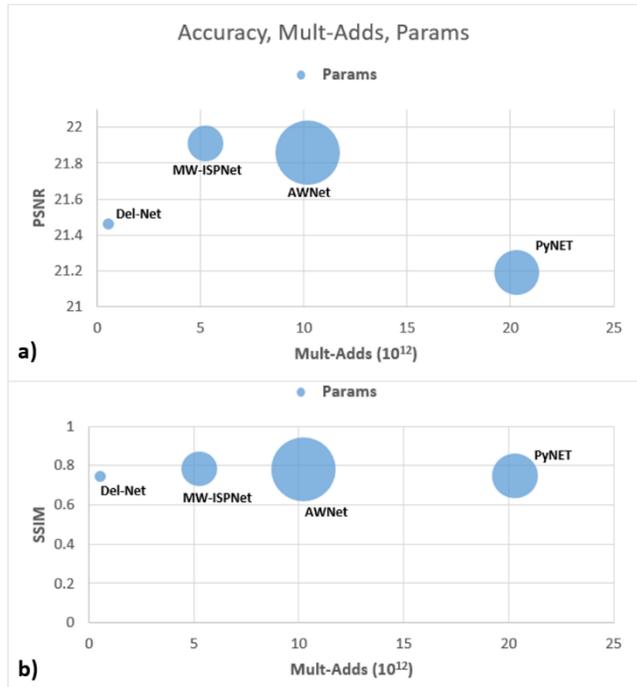

**Figure 4:** Trade-off between performance vs. number of Mult-Add operations and parameters. The size of the circle represents the number of parameters (in $10^6$). The x-axis denotes Mult-Adds (in $10^{12}$). In a) and b) the y-axis denotes PSNR and SSIM respectively.

## 5.2 Qualitative Evaluation

The visual evaluation of Del-Net is shown in Fig. 5, Fig. 6, Fig. 7, and Fig. 8. In Fig. 5, the image patches demonstrate the colour fidelity of Del-Net compared to the state-of-the-art networks. Del-Net reproduces colour better than PyNET and is of comparable quality to MW-ISPNet's results. AWNet's images have a darker shade in most cases. In Fig. 6, we provide patches to compare sharpness of results, with areas to focus on outlined in red. The detail recovery capability of Del-Net is inferior to that of MW-ISPNet, but comparable to that of PyNET despite extremely low Mult-Add operations. Fig. 7 shows the areas where MW-ISPNet is noisy, whereas Del-Net achieves good trade-off between denoising and detail retention. MW-ISPNet further has noise in low-light areas, which is not observed in PyNET, AWNet, and Del-Net. In Fig. 8, it can be seen that MW-ISPNet oversharpens some areas thus resulting in artifacts in grill areas as well as on the yellow wall. We conclude that for the considerably low Mult-Adds that Del-Net has, in majority of the cases it has better detail retention compared to PyNET, better denoising compared to MW-ISPNet, and better colour enhancement compared to AWNet.

## 6 ABLATION STUDY

In the ablation study, we demonstrate the effectiveness of the SCA and EAM modules used in Del-Net. In one experiment, we remove the SCA modules, in the other we remove the EAM blocks, and finally we remove both the modules from Del-Net.

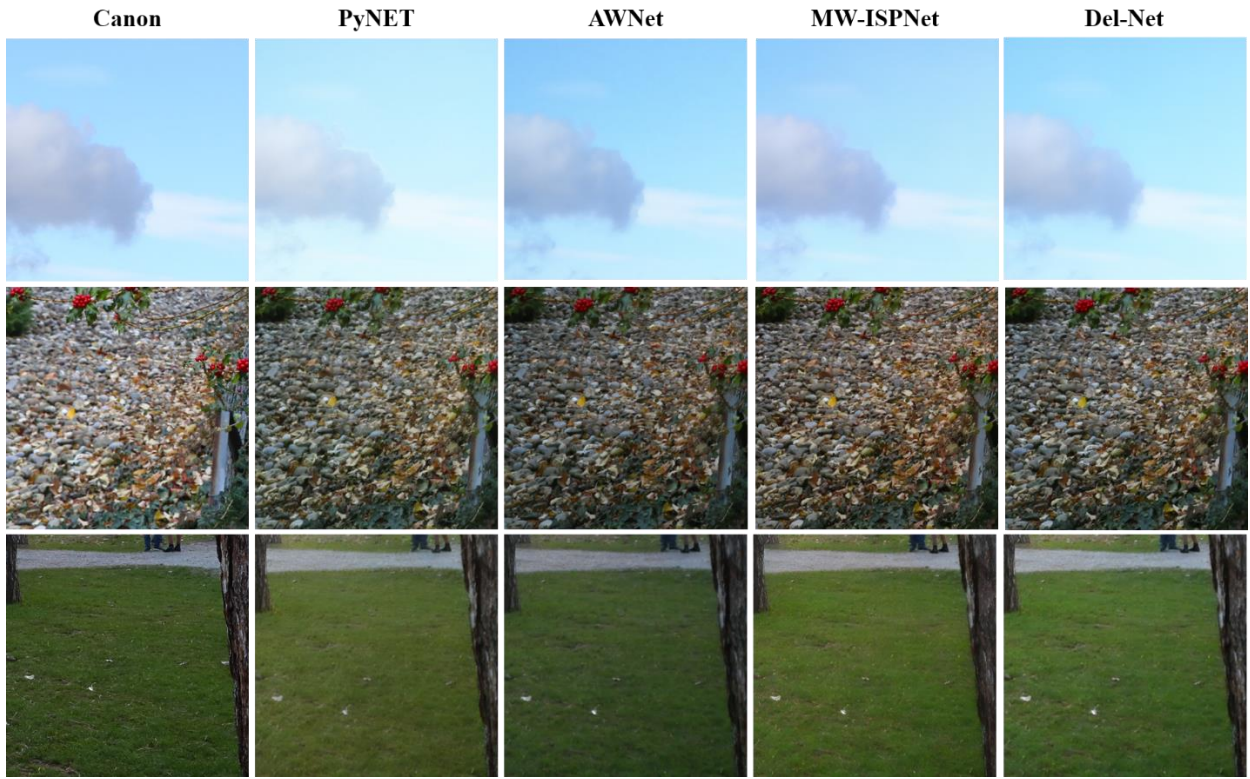

**Figure 5:** Comparison of colour enhancement capability.

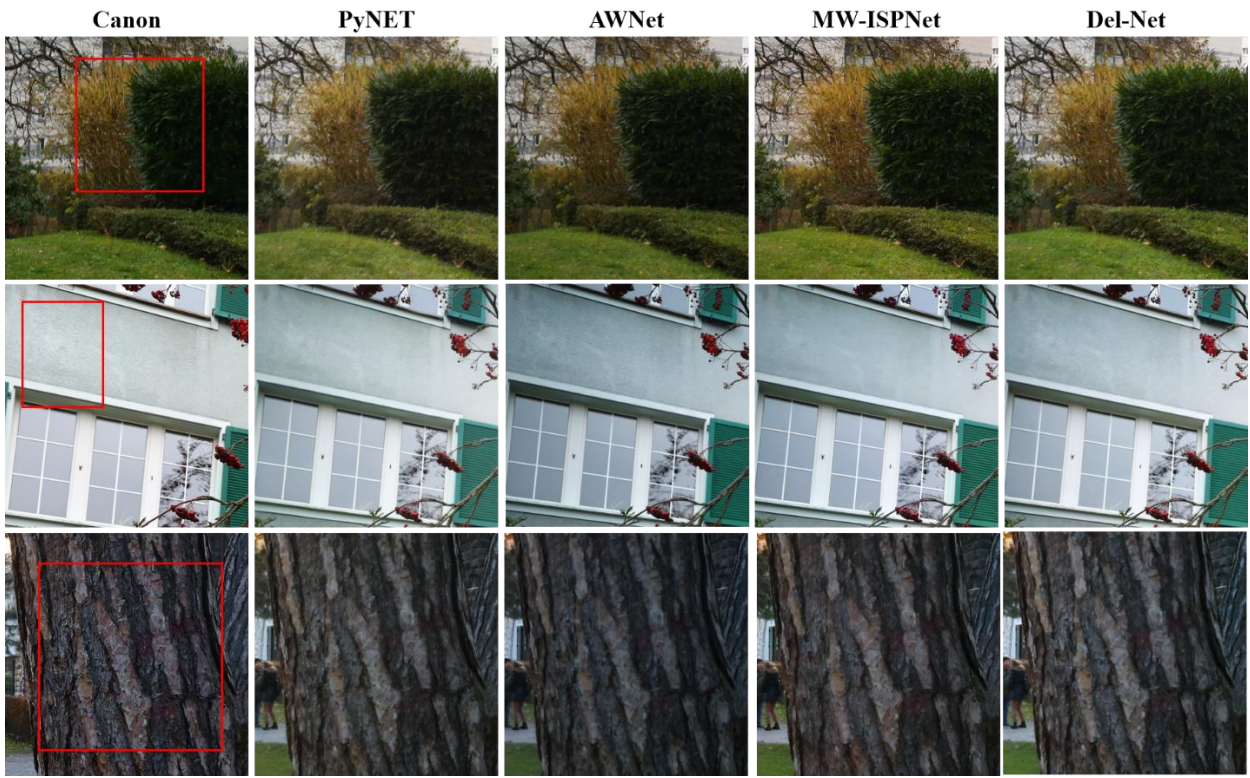

**Figure 6:** Comparison of detail retention capability. Zoom in for better views.

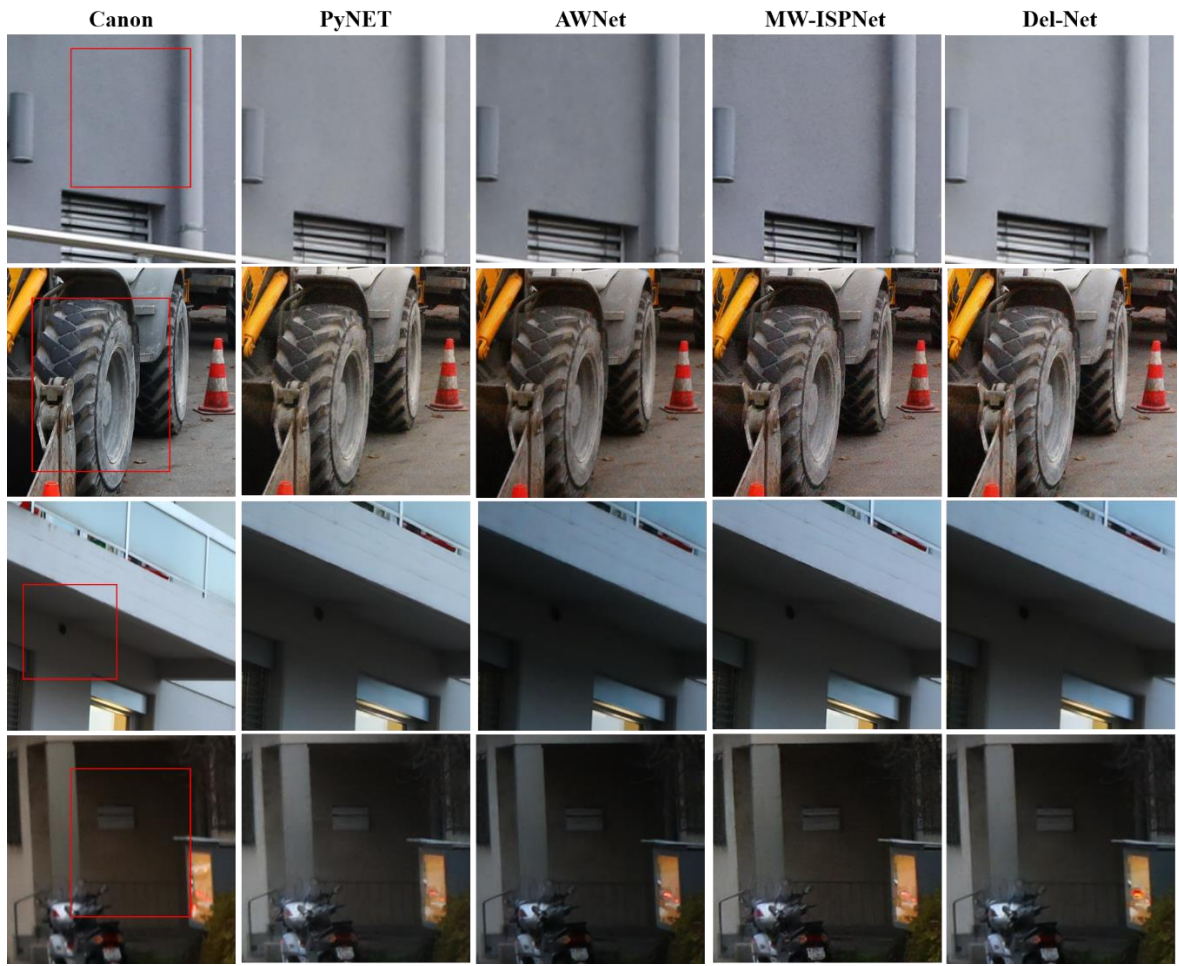

**Figure 7:** Comparison of denoising capability. Zoom in for better views.

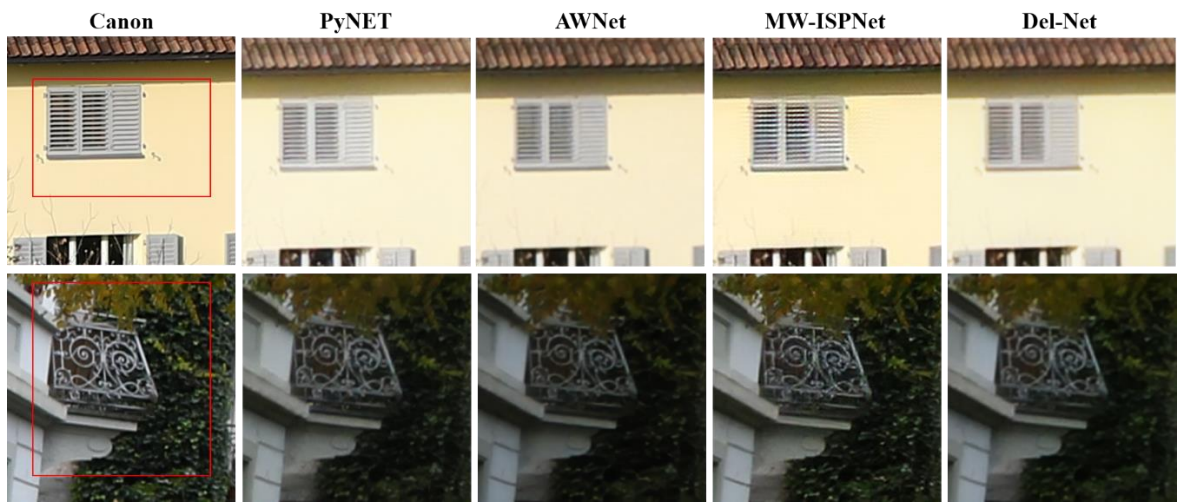

**Figure 8:** Comparison showing artifacts. Zoom in for better views.

The PSNR, SSIM, and CIE2000 numbers for these experiments on the same Zurich dataset are shown in Table III. Visual comparison has been done in Fig. 9. The addition of the EAM blocks improved the denoising strength of the network, resulting in an increased PSNR value. It is also evident that addition of the SCA blocks has led to better CIE2000 value which is in accordance with the fact that the SCA blocks helps in selecting the most useful global features and discarding the less relevant ones, thus helping in colour enhancement.

**TABLE III.** EFFECTS OF SCA AND EAM BLOCKS

| Network | SCA blocks | EAM blocks | PSNR | SSIM | CIE2000 |
|---|---|---|---|---|---|
| UNet | | | 20.44 | 0.709 | 10.51 |
| UNet+SCA | ✓ | | 20.55 | 0.719 | 10.09 |
| UNet+EAM | | ✓ | 20.82 | 0.715 | 10.24 |
| Del-Net | ✓ | ✓ | **21.46** | **0.745** | **9.64** |

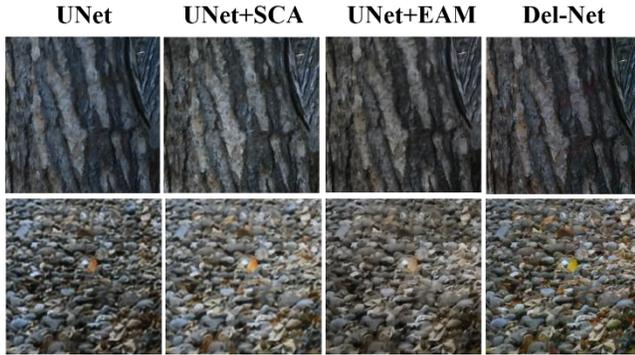

**Figure 9:** Visual comparison of ablation study experiments.

## 7 CONCLUSION

In this paper, we propose Del-Net – a single end-to-end deep learning model – to learn the entire ISP pipeline to convert raw bayer data to high quality sRGB image within reasonable complexity suitable for smartphone deployment. Del-Net is a multi-scale architecture that uses a combination of SCA and EAM blocks. SCA blocks help to capture global details both channel-wise and spatial-wise, while EAM blocks helps in denoising. In our experiments, Del-Net produces comparable visual reconstruction quality as compared to state of the art networks with a reduction of almost 90% in Mult-Adds on publicly available Zurich dataset, thus making it suitable for smartphone deployment.